\documentclass[12pt]{article}
\usepackage[latin1]{inputenc}
\usepackage[spanish]{babel}
\usepackage{amsmath}

\begin{document}
\vskip 1truein
\begin{center}
{\Large {\bf Teor\'{\i}a Autodual de Esp\'{\i}n 2 revisada}}
\vskip 5pt
{\bf P\'{\i}o J. Arias}${}^{a, c, }${\footnote {e-mail:
parias@fisica.ciens.ucv.ve}}{\bf y} {\bf Rolando Gaitan D.}${}^{b,
}${\footnote {e-mail: rgaitan@fisica.ciens.ucv.ve}}${}^{,
}${\footnote {Trabajo presentado en el IV Congreso de la Soc. Ven.
de F\'{\i}sica,
Margarita, Nov. 2003}}\\
${}^a${\it Centro de F\'{\i}sica Te\'orica y Computacional,
Facultad de Ciencias, Universidad Central de Venezuela, AP 47270,
Caracas
1041-A, Venezuela.}\\
${}^b${\it Grupo de F\'{\i}sica Te\'orica, Departamento de F\'\i
sica, Facultad de Ciencias y Tecnolog\'\i a, Universidad de
Carabobo, A.P. 129
Valencia 2001, Edo. Carabobo, Venezuela.}\\
${}^c${\it Centro de Astrof\'{\i}sica Te\'orica, Facultad de
Ciencias, Universidad de Los Andes, La Hechicera, M\'erida 5101,
Venezuela}
\end{center}
\vskip .1truein
\begin{abstract}

Se estudian los v\'{\i}nculos lagrangianos  de la teor\'{\i}a
autodual de esp\'{\i}n 2 en un espacio-tiempo plano $2+1$
dimensional y la acci\'on reducida de un grado de libertad es
obtenida. Partiendo de esta formulaci\'on se calcula el \'algebra
de operadores mec\'anico-cu\'anticos y se explora la
contribuci\'on del esp\'{\i}n en los generadores de
transformaciones.

\vskip .1truein

{\it palabras clave}: Formulaci\'on Lagrangiana de campos,
\'algebra de Poincar\'e

\vskip .1truein

Lagrangian constraints of the spin 2 selfdual theory in a $2+1$
flat space-time are studied and the one degree of freedom reduced
action is obtained. From this formulation, the quantum operator
algebra is computed and the spin contribution on transformation
generators is explored.

\vskip .1truein

{\it keywords }: Field's Lagrangian formulation, Poincar\'e
algebra

{\it pacs numbers}: $11.10.Ef$, $03.70.+k$, $11.30.Cp$

\end{abstract}
\vskip .1truein

\section{Introducci\'on}

Las teor\'ias autoduales en dimensiones impares han recibido
atenci\'on desde hace tiempo [1]. Particularmente, en dimensi\'on
$2+1$  poseen un inter\'es inspirado fundamentalmente por su
relación con la f\'{\i}sica de altas temperaturas en $3+1$
dimensiones [2] y con la f\'{\i}sica de la materia condensada
[3,4].

En la siguiente secci\'on estudiamos los v\'{\i}nculos
Lagrangianos de la teor\'{\i}a autodual de spin 2 y
construimos la acci\'on
reducida que permite establecer de manera inmediata el \'algebra
de operadores sin pasar por el procedimiento can\'onico de Dirac.
En la secci\'on 3 discutimos el esp\'{\i}n de la excitaci\'on via
los generadores del grupo de Poincar\'e.

\section{La acci\'on reducida}

La acci\'on de la teor\'{\i}a autodual del campo $h_{\mu \nu }$
con esp\'{\i}n 2 en un espacio plano es [5]
\begin{equation}
S_{sd}=\frac m2 \int d^3 x (\epsilon ^{\mu \nu \lambda }{h_\mu
}^\alpha
\partial _\nu
h_{\lambda \alpha }-m(h_{\mu \nu }h^{\nu \mu }-h^2 )) \, \, ,
\label{eq1}
\end{equation}
donde $h$ es la traza del campo autodual, $\epsilon ^{012} \equiv
\epsilon ^{12} =+1$ y m\'etrica $\eta = diag(-1,+1,+1)$. Por ser
una teor\'{\i}a de primer orden, la ecuaci\'on de movimiento
proveniente de la extremal de $S_{sd}$ constituye los nueve
v\'{\i}nculos Lagrangianos primarios
\begin{equation}
E^{\mu \rho }\equiv \epsilon ^{\mu \nu \lambda }\partial _\nu
{h_\lambda}^\rho + m(\eta ^{\mu \rho }h - h^{\rho \mu })  \approx
0 \, \, . \label{eq2}
\end{equation}
La preservaci\'on de \'estos proporciona las aceleraciones $
{\ddot h}_{k \rho }$ m\'as tres v\'{\i}nculos secundarios, que
pueden ser escritos en la capa de masa como
\begin{equation}
\partial _\mu E^{\mu \rho } - m\epsilon ^{\rho \mu \alpha }
E_{\mu \alpha  } \equiv -m^2 \epsilon ^{\rho \mu \alpha } h_{\mu
\alpha }
\approx 0
\, \, . \label{eq3}
\end{equation}
Siguiendo con el procedimiento, \'estos conducen a tres nuevos
v\'{\i}nculos,
\begin{equation}
-m^2 \epsilon ^{\rho \mu \alpha } {\dot h}_{\mu \alpha }
\approx 0
\, \, . \label{eq4}
\end{equation}
Obs\'ervese que la combinaci\'on sobre la capa de masas $-m^2
\epsilon ^{\rho \mu \alpha } \partial _\rho h_{\mu \alpha }
\approx 2 m^3 h \approx 0 $ es un v\'{\i}nculo, con lo cual la
preservaci\'on de (4) proporciona las aceleraciones $ {\ddot h}_{o
k }$ y el \'ultimo v\'{\i}nculo,
\begin{equation}
2 m^3 \dot h \approx 0 \, \, , \label{eq5}
\end{equation}
que al ser preservado permite despejar la aceleraci\'on faltante $
{\ddot h}_{o o }$, culminando el proceso. Entonces se tiene un
sistema de $9+3+3+1=16$ v\'{\i}nculos, indicando que hay una sola
excitaci\'on.

Si el v\'{\i}nculo secundario es reescrito en la capa de masas
como $\partial _\mu E^{\mu \rho } \equiv m
\partial ^\rho h - m\partial _\mu h^{\rho \mu }
\approx 0$, se puede ver f\'acilmente que el sistema de v\'{\i}nculos 
Lagrangianos
equivale a describir un campo sim\'etrico, transverso y sin traza
que satisface la ecuaci\'on $(\partial ^2 - m^2){h^{Tt}}_{\mu
\nu}=0$, donde
$\partial ^2 \equiv \partial _{\mu}\partial ^{\mu}=-\partial _o\partial _o + 
\Delta$.

Para exponer a nivel Lagrangiano la \'unica excitaci\'on de la
teor\'{\i}a, consideramos la descomposici\'on $2+1$ [6]:
\begin{eqnarray}
n=h_{00} \, \, ,
\,\,N_i=h_{i0} \, \, ,
\,\,M_i=h_{0i} \, \, ,\nonumber \\
H_{ij}=\frac{1}{2}(h_{ij}+h_{ji}) \, \, ,
\,\,V=\frac{1}{2}\,\epsilon _{ij}h_{ij} \, \, , \label{eq6}
\end{eqnarray}
con componentes transverso-longitudinales dadas por
\begin{eqnarray}
N_i \equiv \epsilon _{ik}\partial _k N^T + \partial _i N^L \, \, ,
\,\,M_i \equiv \epsilon _{ik}\partial _k M^T + \partial _i M^L \,
\, , \nonumber \\
H_{ij} \equiv (\delta _{ij}\Delta -\partial _i \partial _j)H^T
+\partial _i \partial _j H^L  +( \epsilon _{ik}\partial _k
\partial _j+\epsilon _{jk}\partial _k \partial _i )H^{TL} \, \, ,
\label{eq7}
\end{eqnarray}
en la acci\'on (1). Usando las ecuaciones de movimiento que se
obtienen es posible reescribir la acci\'on autodual en la forma
reducida
\begin{eqnarray}
S^*_{sd}=\int d^3 x \{P\dot{Q}-\frac12 P^2 + \frac12 \,Q(\Delta
-m^2)Q\}
\, \, ,
\label{eq8}
\end{eqnarray}
donde se han definido las variables $Q\equiv \sqrt 2 \, \Delta
H^{T}$ y $ P\equiv \sqrt 2 \,m \Delta H^{TL}$, que satisfacen
$P=\dot{Q}$. Si promovemos los campos $h_{\mu\nu}$ a sus
operadores mec\'anico-cu\'anticos, se puede obtener el \'algebra
de \'estos partiendo de la regla fundamental
\begin{eqnarray}
\big[Q(x),P(y) \big]_{x^o=y^o}=i{\delta}^2(\vec{x}-\vec{y})\, \, ,
\label{eq9}
\end{eqnarray}
con la ayuda de la descomposici\'on $2+1$ transverso-longitudinal.
Utilizando propiedades del s\'{\i}mbolo de Levi-civita, los
conmutadores no nulos que se obtienen son
\begin{eqnarray}
\big[h_{io}(x),h_{jk}(y) \big]
=\big[h_{oi}(x),h_{jk}(y)\big]=\frac{i}{2m^2}\{
{p^{(m)}}_{ij}\partial _k+{p^{(m)}}_{ik}
\partial _j-{p^{(m)}}_{jk}\partial _i \}{\delta}^2(\vec{x}-
\vec{y})\, \, , \label{eq10}
\end{eqnarray}
\begin{eqnarray}
\big[h_{io}(x),h_{oo}(y)
\big]=\big[h_{oi}(x),h_{oo}(y)\big]=-\frac{i}{2m^4}\,\partial
_i\Delta {\delta}^2(\vec{x}-\vec{y})
\, \, ,
\label{eq11}
\end{eqnarray}
\begin{eqnarray}
\big[h_{io}(x),h_{ok}(y)\big]=\big[h_{io}(x),h_{ko}(y)\big]=\big[h_{oi}(x),h_{ko}(y)\big]
=\frac{i\,\epsilon _{ki}}{2m^3}\,\Delta
{\delta}^2(\vec{x}-\vec{y})
\, \, ,
\label{eq12}
\end{eqnarray}
\begin{eqnarray}
\big[h_{oo}(x),h_{ij}(y) \big]=\frac{i}{2m}\{ \epsilon
_{ki}\,{p^{(m)}}_{kj}+\epsilon _{kj}\,{p^{(m)}}_{ki}\}
{\delta}^2(\vec{x}-\vec{y})
\, \, ,
\label{eq13}
\end{eqnarray}
\begin{eqnarray}
\big[h_{ij}(x),h_{kl}(y)\big] =\frac{i}{4m}\{\epsilon
_{ik}\,{p^{(m)}}_{jl} +  \epsilon _{jk}\,{p^{(m)}}_{il}
+  \epsilon _{il}\,{p^{(m)}}_{jk}+\epsilon
_{jl}\,{p^{(m)}}_{ik}\}{\delta}^2(\vec{x}-\vec{y})\, \, ,
\label{eq14}
\end{eqnarray}
donde ${p^{(m)}}_{ij}= \delta _{ij} - \frac{\partial _i
\partial _j}{m^2}$ es el proyector transversal en la capa $\Delta =m^2$.

Es de esperarse que el \'algebra de operadores obtenida mediante
el procedimiento de la acci\'on reducida sea equivalente al de la
realizaci\'on a la Dirac[7], hecho que est\'a sustentado por un
teorema[8] que garantiza la igualdad entre los corchetes de Dirac
y los de Poisson calculados con las variables reducidas.

\section{Generadores del \'algebra de Poincar\'e}

Con la finalidad de construir los generadores del \'algebra de
Poincar\'e, uno puede determinar el tensor momento-energ\'ia
sim\'etrico del campo autodual ($T^{\alpha \beta}$)
extendendiendo la
acci\'on autodual (1) al caso de un espacio-tiempo dotado con una
  m\'etrica general $g_{\mu \nu}$, con lo cual
\begin{eqnarray}
T^{\alpha \beta} =\bigg[\frac{2}{\sqrt{-g}}\, \frac{\delta
S}{\delta g_{\alpha \beta}}\bigg]_{g_{\mu \nu}=\eta _{\mu \nu}} =
\frac{m^2}{2}\,\big( h^{\sigma
\alpha}{h^{\beta}}_{\sigma}+h^{\sigma
\beta}{h^{\alpha}}_{\sigma}-h\,h^{\alpha \beta}-h\,h^{\beta
\alpha}  \nonumber \\-\,\eta ^{\alpha \beta}h_{\mu \nu}h^{\nu \mu}
+ \eta ^{\alpha \beta}h^2 \big)  -\frac{m}{2}\,\big(
\partial _{\sigma} t^{\alpha \beta \sigma}+
{h_{\sigma}}^{\alpha}E^{\sigma \beta} +
{h_{\sigma}}^{\beta}E^{\sigma \alpha}\big) \, \,,\label{eq15}
\end{eqnarray}
donde $t^{\alpha \beta \sigma}\equiv  \epsilon ^{\mu \alpha \nu
}{h_\mu }^\beta {h_\nu }^\sigma  + \epsilon ^{\mu \beta \nu
}{h_\mu }^\alpha {h_\nu }^\sigma $ y $E^{\mu \rho }\equiv \epsilon
^{\mu \nu \lambda }\partial _\nu {h_\lambda}^\rho + m(\eta ^{\mu
\rho }h - h^{\rho \mu })$, como en (2).

Los generadores de translaciones $\mathcal{P}^{\mu}= \int d^2
x\,T^{o \mu}$ se expresan en t\'erminos de $Q$ y su momento
conjugado, observ\'andose que coinciden con los de un campo
escalar, es decir
\begin{eqnarray}
\mathcal{P}^o=\frac{1}{2}\int d^2x \{ P^2 +\partial _i Q
\partial _i Q +m^2 Q^2 \} \, \,,\label{eq16}
\end{eqnarray}
\begin{eqnarray}
\mathcal{P}^i=-\int d^2x P \partial _i Q  \, \,.\label{eq17}
\end{eqnarray}
De igual manera ocurre con los generadores de rotaciones
$\mathcal{J}^{ij}=\int d^2x \{ x^i T^{0j}-x^j T^{0i} \}\equiv
\epsilon ^{ij} \mathcal{J}$, donde
\begin{eqnarray}
\mathcal{J}=- \int d^2x P\epsilon ^{kl}x^k \partial _l Q \,
\,,\label{eq18}
\end{eqnarray}
ya que en dos dimensiones ellas estan descritas por el grupo
$O(2)$. Pero la contribuci\'on expl\'{\i}cita del esp\'{\i}n se
pone de manifiesto cuando escribimos los generadores de los {\it
boosts} de Lorentz
\begin{eqnarray}
\mathcal{J}^{i0} =\frac{1}{2}\int d^2x \, x^i \{P^2 +\partial _i Q
\partial _i Q +m^2 Q^2 \} -x^0 \mathcal{P}^i + 2m \int d^2x \, P
\frac{\epsilon ^{ij} \partial _j}{\Delta}\, Q
\, \,,
\label{eq19}
\end{eqnarray}
donde se observa el t\'{\i}pico factor 2 de esp\'{\i}n en el
t\'ermino ''singular infrarrojo'', indicando que $Q$ no transforma
como un campo escalar, como debe esperarse.

Para remover la ''singularidad infrarroja'', se expande $Q$ en
ondas planas
\begin{eqnarray}
Q(x)=\int \frac{d^2 k}{2\pi \sqrt{2w(\bf{k})}} \, \{ e^{-ik_{\mu}
x^{\mu}}a({\bf{k}}) + e^{ik_{\mu} x^{\mu}}a^+(\bf{k}) \}
,
\label{eq20}
\end{eqnarray}
con $k^o =w(\bf{k}), \overrightarrow{k}=\bf{k}$\, y
\,$[a({\bf{k}}),a^+({\bf{k'}})]=\delta ^2(\bf{k}-\bf{k'})$.
Entonces, los generadores de translaciones y rotaciones son
representados por
\begin{eqnarray}
\mathcal{P}^\mu=\int d^2k\,  k^\mu a^+({\bf{k}})a({\bf{k}}) \, \,,
\label{eq21}
\end{eqnarray}
\begin{eqnarray}
\mathcal{J}^{ij}=\int d^2k\, a^+({\bf{k}})\frac{\epsilon
^{ij}}{i}\,\frac{\partial}{\partial \theta }\, a({\bf{k}})\, \,,
\label{eq22}
\end{eqnarray}
con $\tan{\theta}=k_2/k_1$. En esta
representaci\'on, el generador de {\it boosts} exhibe la
''singularidad infrarroja''
\begin{eqnarray}
\mathcal{J}^{i0}=\frac{i}{2}\,\int d^2 k \,w({\bf{k}})\{
a^+({\bf{k}})\overline{\partial _i}\, a({\bf{k}})\} -2m\int d^2 k
\,\frac{\epsilon ^{ij}\,k^j}{{\bf{k}}^2} \,
a^+({\bf{k}})a({\bf{k}})
\, \,,
\label{eq23}
\end{eqnarray}
donde $a^+({\bf{k}})\overline{\partial _i}\, a({\bf{k}})\equiv
a^+({\bf{k}})\partial _i\, a({\bf{k}}) -a({\bf{k}})\partial _i\,
a^+({\bf{k}})$.

Seguidamente, se realiza una transformaci\'on de fase [9] de la
forma $a({\bf{k}})\longrightarrow
e^{i\,s\,\frac{m}{|m|}\,\theta}a({\bf{k}})$  sobre los operadores
creaci\'on-aniquilaci\'on, con lo cual los generadores de
translaciones no son afectados, mientras que los de rotaciones y
{\it boosts} son ahora
\begin{eqnarray}
\mathcal{J}^{ij}=\int d^2 k \, a^+({\bf{k}})\frac{\epsilon
^{ij}}{i}\,\frac{\partial}{\partial \theta }\,
a({\bf{k}})+s\frac{m}{|m|}\,\int d^2 k \,\epsilon
^{ij}\,a^+({\bf{k}})a({\bf{k}})
\, \,,
\label{eq24}
\end{eqnarray}
\begin{eqnarray}
\mathcal{J}^{i0}=\frac{i}{2}\,\int d^2 k \,w({\bf{k}})\{
a^+({\bf{k}})\overline{\partial _i}\,
a({\bf{k}})\}+2\frac{m}{|m|}\int d^2 k \,\frac{\epsilon
^{ij}k^j}{w({\bf{k}})+|m|} \,a^+({\bf{k}})a({\bf{k}})\nonumber \\
+(s-2)\,\frac{m}{|m|}\,\int d^2 k \,w({\bf{k}})\,\frac{\epsilon
^{ij}k^j}{{\bf{k}}^2} \,a^+({\bf{k}})a({\bf{k}}) \, \,.
\label{eq25}
\end{eqnarray}
Inmediatamente vemos que $s=2$ remueve la singularidad y que
cualquier otro valor diferente asignado a este par\'ametro
mantiene el comportamiento singular de los generadores. N\'otese
adem\'as que el valor de esp\'{\i }n $2\frac{m}{|m|}$ es
recuperado, cuya sensibilidad al cambio de signo de $m$ refleja la
helicidad del grado de libertad que se propaga. Desde el punto de
vista Lagrangiano, esta manifestaci\'on de la helicidad proviene
del signo del t\'ermino lineal en $m$ de la acci\'on (1), y sea
cual sea \'este debe mantenerse el signo presentado en el
t\'ermino cuadr\'atico en $m$ para poder obtener un Hamiltoniano
positivo-definido, como el que se deduce de la transformada de
Legendre de la acci\'on reducida (8).

Finalmente, puede mostrarse que el \'algebra de Poincar\'e es
satisfecha durante todo el proceso, antes y despu\'es de
realizarse la mencionada transformaci\'on de fase, siempre y
cuando se defina de manera adecuada el contorno de integraci\'on
alrrededor de la ''singularidad infrarroja''.

\section{Conclusi\'on}

En el estudio de la teor\'{\i}a del campo autodual hemos mostrado
que la formulaci\'on de acci\'on reducida constituye una
herramienta \'util para la construcci\'on de la teor\'{\i}a
cu\'antica correspondiente, evitando el procedimiento extenuante
de cuantizaci\'on a la Dirac. As\'{\i} mismo, este formalismo que
describe una exitaci\'on masiva para el caso estudiado, permite
precisar la contribuci\'on del esp\'{\i}n estableciendo el
comportamiento no escalar del \'unico grado de libertad.
All\'{\i}, hemos observado que es posible evitar la ''singularidad
infrarroja'' mediante una transformaci\'on de fase \'unica sobre
los operadores creaci\'on-aniquilaci\'on y que \'algebra de
Poincar\'e es satisfecha.

\vspace{1pt}
{\bf Agradecimientos}

Los autores agradecen a asistencia t\'ecnica de Orlando Alvarez
LLamoza, as\'{\i} como el apoyo recibido por parte del Decanato de
la FACYT-UC y la OPSU. Este trabajo es parcialmente apoyado por el
proyecto G-2001000712 del FONACIT.

\end{document}